\documentclass[twocolumn]{aastex631}

\newcommand{\revI}[1]{\textrm{#1}}
\newcommand{\revII}[1]{\textrm{#1}}
\usepackage{bm}
\usepackage{csquotes}
\usepackage{subfiles}
\usepackage{amsmath}
\usepackage{comment}

\makeatletter
\def\footnoterule{\kern-3\p@
  \hrule \@width 2in \kern 2.6\p@} 
\makeatother

\makeatletter
\patchcmd{\ltx@foottext}{%
  .5\textwidth\advance\hsize-18pt}{%
  \linewidth\advance\hsize-1.8em%
}{}{}
\makeatother

\begin{document}

\title{The Role of the Convective Kissing Instability in Cataclysmic Variable Evolution}

\author[0000-0002-4524-9497]{Conor M. Larsen}
\affiliation{Department of Physics and Astronomy,
University of Delaware, Newark, DE, 19716, USA}\email{cmlarsen@udel.edu}

\author[0000-0002-0506-5124]{James MacDonald}
\affiliation{Department of Physics and Astronomy,
University of Delaware, Newark, DE, 19716, USA}

\begin{abstract}

The convective kissing instability (CKI) \revI{is postulated to occur} in low mass stars around the fully convective transition. Non-equilibrium $^{3}$He burning leads to the merging of core and envelope convective zones, which causes abrupt decreases in the stellar radius. It has been suggested \revI{by} \cite{CKI_initial} that these effects may be relevant for cataclysmic variables (CVs). We \revI{have} performed stellar evolution \revI{modeling} to study the role of the CKI in CV evolution. We find that the CKI has no effect on normal CVs which evolve via magnetic braking and gravitational radiation above the period gap. CKI cycles either do not occur or are abruptly halted once mass transfer begins. If only gravitational radiation is considered, the CKI does occur. The abrupt radius changes can cause detachment phases which produce small period gaps \revI{with widths of} a few minutes. We describe how the size of the period gaps is controlled by the $^{3}$He profiles of the secondaries. We also discuss how the results of this study apply to the evolution of strong field polars, where the magnetic field of the white dwarf is strong enough to suppress magnetic braking.

\end{abstract}


\section{Introduction} \label{sec:intro}
Cataclysmic variables (CVs) are close binary stars containing a white dwarf (WD) and \revI{typically} a low mass, main sequence (MS) secondary which is actively transferring mass to the WD \cite[see][for a review]{Warner_text}. The evolution of CVs is an active research area. The main observed feature to be captured in evolutionary tracks is the CV orbital period gap, which is the lack of CVs found with orbital periods between ${\sim}2-3$ hours \citep{Period_gap}. CVs above the gap experience angular momentum loss (AML) through gravitational radiation and magnetic braking from the secondary's stellar wind. The upper edge of the period gap corresponds to the fully convective transition in low mass stars. In the classical explanation of CV evolution, known as the disrupted magnetic braking model \citep{RVJ}, magnetic braking ceases when the star becomes fully convective. The high mass transfer rates above the period gap drive the secondaries out of thermal equilibrium, causing them to be bloated in size. When the AML rate abruptly drops, the mass transfer rate also drops, allowing the star to relax into equilibrium. The star shrinks inside its Roche lobe, cutting of mass transfer. 

While the disrupted magnetic braking model can reproduce the period gap, the idea has been questioned due to mounting evidence that fully convective stars can maintain strong magnetic fields \citep{full_conv_B_field_indirect, fully_conv_B_field}. Since entering the gap requires the secondaries to become fully convective, understanding the CV period gap requires an understanding of the stellar response to the \revI{fully} convective transition. While the current consensus still leans to a disrupted AML rate to explain the period gap \citep[e.g.][]{SBD_model_for_CVs}, it is also important to test other ideas related to the fully convective transition. 

The convective kissing instability (CKI), \revI{discovered} by \cite{CKI_initial}, occurs in \revI{1D stellar evolution models} with masses around the fully convective transition ($M\sim0.35$ $M_{\odot}$). The instability is driven by non-equilibrium $^{3}$He burning which causes mergers between convective zones. These mergers in turn cause abrupt changes in the radius and luminosity, which are sometimes refereed to as \enquote{glitches} (a more detailed overview of the CKI is given in the following section). Interest in the CKI mostly revolves around the \textit{Gaia} M-dwarf gap, which is a low density region in the MS of the \textit{Gaia} DR2 color magnitude diagram (CMD) at $M_{G}\approx10$ \citep{M_dwarf_gap}. Theoretical evolutionary tracks confirm that non-equilibrium $^{3}$He burning and the subsequent merging of convective zones are the cause of the \textit{Gaia} gap \citep{MacDonald_Gizis_2018,Baraffe_Chabrier_2018,Mansfield_Kroupa_2021}. Population synthesis performed by \cite{Feiden_Skidmore_Jao_2021} is able to generally reproduce the \textit{Gaia} gap, although the exact location of the synthetic gap in the \textit{Gaia} CMD differs from the observed.

In the discovery paper, \cite{CKI_initial} conjectured that the CKI may be important for CVs, suggesting that the radius \enquote{glitches} could lead to detachment or enhanced mass loss rates. Here we present detailed stellar evolution models of CVs to demonstrate the role of the CKI in CV evolution. This paper includes an overview of the CKI (section \ref{sec:CKI}), a description of the stellar evolution models (section \ref{sec:models}), the results of the models (section \ref{sec:results}), a discussion of the results (section \ref{sec:discussion}) and a conclusion (section \ref{sec:conclusion}).

\section{Overview of the Convective Kissing Instability} \label{sec:CKI}

\begin{figure}
    \centering    
    \includegraphics[width=0.45\textwidth]{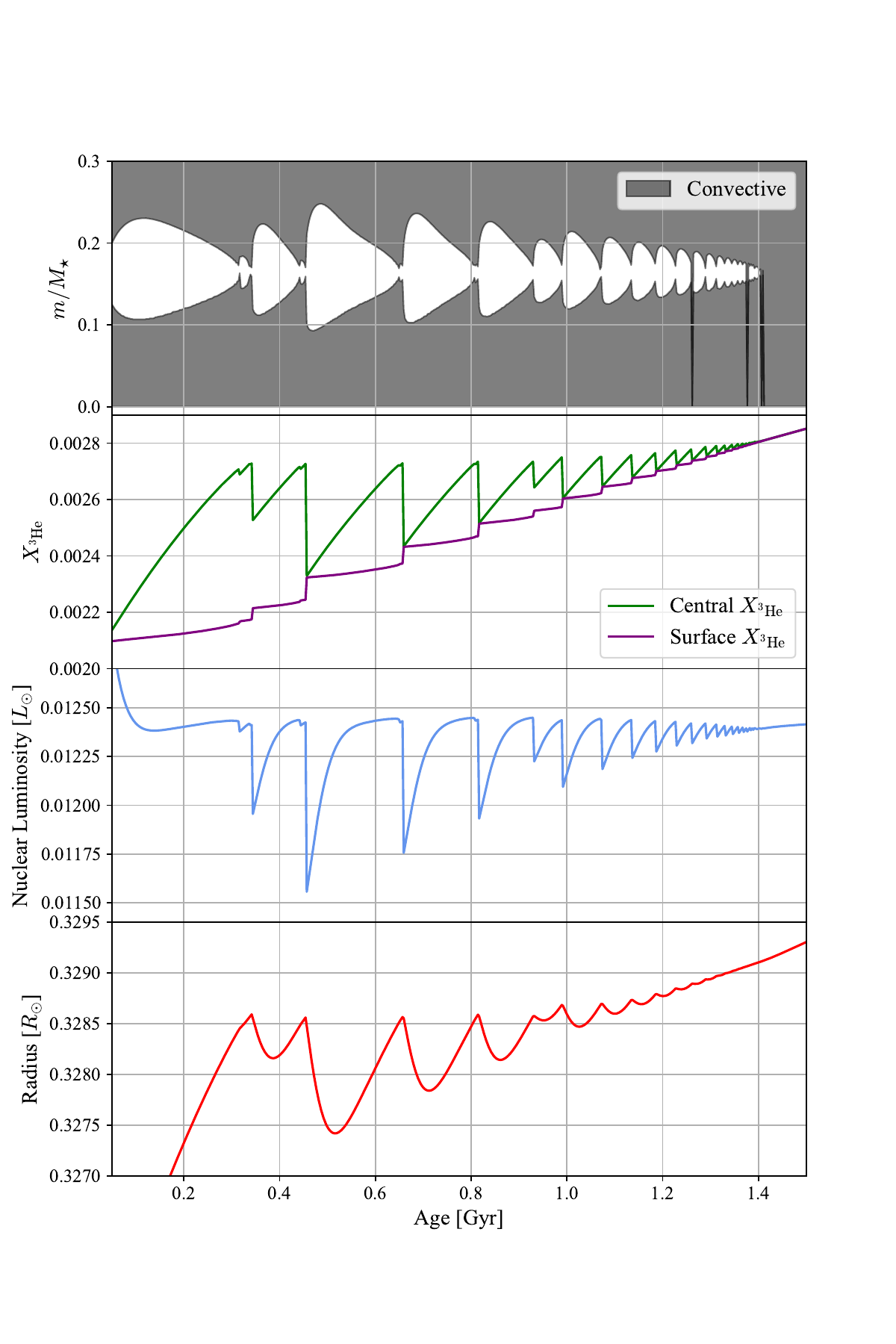}
    \caption{The convective kissing instability in a 0.335 $M_{\odot}$ model with $Z = 0.02$. From top to bottom, the plots are (1) the mass coordinate of the convective (grey) and radiative (white) zones vs. time, (2) the central and surface $^{3}$He mass fractions vs time, (3) the nuclear luminosity vs. time and (4) the radius vs. time.}
    \label{fig: single cki}
\end{figure}

Figure \ref{fig: single cki} displays the CKI in a 0.335 $M_{\odot}$ model with $Z = 0.02$. The models were produced in the Modules for Experiments in Stellar Astrophysics (\texttt{MESA}), the details of which will be explained next section. The top plot of figure \ref{fig: single cki} shows the merging of convective zones, which is driven by non-equilibrium $^{3}$He burning. The pp I chain consists of the following reactions:

\begin{equation}\label{eq: reaction 1}
    p + p \rightarrow d + e^{+} + \nu_{e},
\end{equation}

\begin{equation}\label{eq: reaction 2}
    p + d \rightarrow {^{3}}\mathrm{He} + \gamma,
\end{equation}

\begin{equation}\label{eq: reaction 3}
    {^{3}}\mathrm{He} + {^{3}}\mathrm{He} \rightarrow {^{4}}\mathrm{He} + 2p.
\end{equation}
Reaction \ref{eq: reaction 2} produces $^{3}$He while reaction \ref{eq: reaction 3} destroys $^{3}$He. At the low central temperatures in models around the fully convective transition ($\lesssim10^{7}$ K), reaction \ref{eq: reaction 2} dominates over reaction \ref{eq: reaction 3} and $^{3}$He is created in the core. The increase in the central $^{3}$He mass fraction ($X_{^{3}\mathrm{He}}$) is shown in the middle-top plot in figure \ref{fig: single cki}. The local increase of $X_{^{3}\mathrm{He}}$ causes a local enhancement of nuclear burning, as reaction \ref{eq: reaction 3} occurs more frequently. This causes the development of a central convective core. 

As the two convective regions are separated by a radiative region, the enhancement of $^{3}$He in the core cannot mix through the rest of the star. As $X_{^{3}\mathrm{He}}$ increases in the core, the nuclear luminosity increases, which pushes the inner convective region outwards. While this occurs, the outer convective region moves inwards until the two convective regions meet. Upon the merging of the convective zones, the entire star is briefly fully convective, allowing $^{3}$He to mix through the entire star, causing a sharp drop in the central $X_{^{3}\mathrm{He}}$ and a sharp rise in the surface $X_{^{3}\mathrm{He}}$. The sharp drop in $X_{^{3}\mathrm{He}}$ in the center causes a sharp drop in the nuclear luminosity, which in turn forces the central region to become radiative. The cycle then repeats as an overabundance of $^{3}$He is again generated in the central region. Over time, the $^{3}$He content of the entire star increases, which damps the CKI and eventually causes the star to become fully convective. Also apparent are smaller pulsations where some $^{3}$He leaks over into the outer convective envelope, but not enough to equalize $X_{^{3}\mathrm{He}}$ in the center and surface, resulting in smaller fluctuations in the nuclear luminosity and radius. These smaller pulsations have been noted in previous works \citep{Mansfield_Kroupa_2021,Mansfield_Kroupa_2023}.

In the context of CVs, the response of the radius to the CKI is important. The drop in nuclear luminosity at each merger causes a sharp drop in the stellar radius. As the secondaries in CVs evolve through this mass range, the sharp drop in radius could cause a detachment. Additionally, the following increase in radius could cause enhanced mass loss rates. \cite{CKI_initial} presented models undergoing constant mass loss rates, finding that for an initial mass of 0.4 $M_{\odot}$, mass loss rates of $\dot{M} = 5 \times10^{-12} \: - \: 1 \times10^{-10}$ $M_{\odot}$/yr, undergo the CKI. For a mass loss rate of $5\times10^{-10}$ $M_{\odot}$ the star is driven far out of thermal equilibrium and the CKI did not occur. In this report, we further expand on the role of the CKI in CVs and mass losing stars by taking into account binary evolution.

\section{Stellar Models} \label{sec:models}
Our stellar evolution models are produced in the Modules for Experiments in Stellar Astrophysics \citep[\texttt{MESA}, version r24.08.1,][]{Paxton2011, Paxton2013, Paxton2015, Paxton2018, Paxton2019, Jermyn2023}. The \texttt{MESA} EOS is a blend of the OPAL \citep{Rogers2002}, SCVH
\citep{Saumon1995}, FreeEOS \citep{Irwin2004}, HELM \citep{Timmes2000},
PC \citep{Potekhin2010}, and Skye \citep{Jermyn2021} EOSes. Radiative opacities are primarily from OPAL \citep{Iglesias1993,
Iglesias1996}, with low-temperature data from \citet{Ferguson2005}
and the high-temperature, Compton-scattering dominated regime by
\citet{Poutanen2017}.  Electron conduction opacities are from
\citet{Cassisi2007} and \citet{Blouin2020}. Nuclear reaction rates are from JINA REACLIB \citep{Cyburt2010}, NACRE \citep{Angulo1999} and
additional tabulated weak reaction rates \citep{Fuller1985, Oda1994,
Langanke2000}.  Screening is included via the prescription of \citet{Chugunov2007}.
Thermal neutrino loss rates are from \citet{Itoh1996}. For our outer boundary condition, we used a combination of COND \citep{COND} and \cite{Castelli_Kurucz_2003} atmosphere models with the surface optical depth at $\tau_{\mathrm{surf}} = 100$. All our models use an initial metallically of $Z = 0.02$ and the initial helium mass fraction is $Y = 0.24 + 2Z$. Convective instability is determined by the Ledoux criterion and the convective energy flux is determined via the mixing length theory of \cite{Cox_Giuli_book}. Throughout we adopt a mixing length alpha of $\alpha_{\mathrm{MLT}} = 2$. Semi-convection is included via the prescription of \cite{semi_conv} with a semi-convection alpha of $\alpha_{\mathrm{sc}} = 0.1$. We also adopt the convective pre-mixing scheme \citep[see][]{Paxton2019}. \revI{Under the convective pre-mixing scheme, convective cells are instantaneously mixed}.

We use the binary module to explore the role of the CKI in CVs. Roche lobe radii in binary systems are computed using the fit of
\citet{Eggleton1983}. Mass transfer rates in Roche lobe
overflowing binary systems are determined following the
prescription of \citet{Ritter1988}. The WD is modeled as a point mass with $M = 0.81$ $M_{\odot}$, which is consistent with the average WD mass in CVs \citep{ave_WD_mass}. Additionally, we assume nova outbursts remove any material transferred by the secondary, therefore the WD mass remains constant throughout the evolution. AML by gravitational radiation is included by the expression of \cite{J_dot_GR}. When included, AML due to magnetic braking is calculated with the \cite{RVJ} expression with $\gamma = 3$.

Resolving the CKI requires the use of a small timestep. Previous studies typically adopt a maximum timestep of $10^{6}$ years or less \citep[e.g.][]{CKI_initial, Mansfield_Kroupa_2023}. \cite{Mansfield_Kroupa_2021} utilized a small maximum timestep of $5\times10^{4}$ years. We find that a maximum timestep of $5\times10^{5}$ years can resolve that CKI at reasonable computation time. To save computation time the maximum timestep would only be applied when the mass drops below $0.39$ $M_{\odot}$ (i.e. when the mass approaches the CKI range).

\section{Results} \label{sec:results}
\subsection{Models with Magnetic Braking}

First we consider models with the \cite{RVJ} magnetic braking law. We do not expect the CKI to occur when strong AML mechanisms act for two reasons: (1) the high AML rates due to magnetic braking cause a high mass loss rate, which drives the secondary out of thermal equilibrium (i.e. the mass loss timescale is smaller than the thermal timescale). \cite{CKI_initial} have demonstrated that the CKI does not occur in models which have been driven out of thermal equilibrium. (2) Since the stars are out of thermal equilibrium, the transition to fully convective occurs at smaller masses, ${\sim}0.2 - 0.25$ $M_{\odot}$. As mentioned in the discussion section of \cite{Baraffe_Chabrier_2018}, the central temperatures at this mass range are too low for the CKI to occur and thus the transition to fully convective should be smooth. 

To verify these claims, we computed models with starting masses of $0.335$ $M_{\odot}$, $0.4$ $M_{\odot}$, $0.6$ $M_{\odot}$ and $0.8$ $M_{\odot}$ all at a starting period 9.5 hours. Magnetic braking was kept on for the entire evolution. Removing magnetic braking at the fully convective transition is already known to cause detachment. The question here is whether the CKI can cause a detachment, or \revI{some} other feature. 

\begin{figure}
    \centering    
    \includegraphics[width=0.45\textwidth]{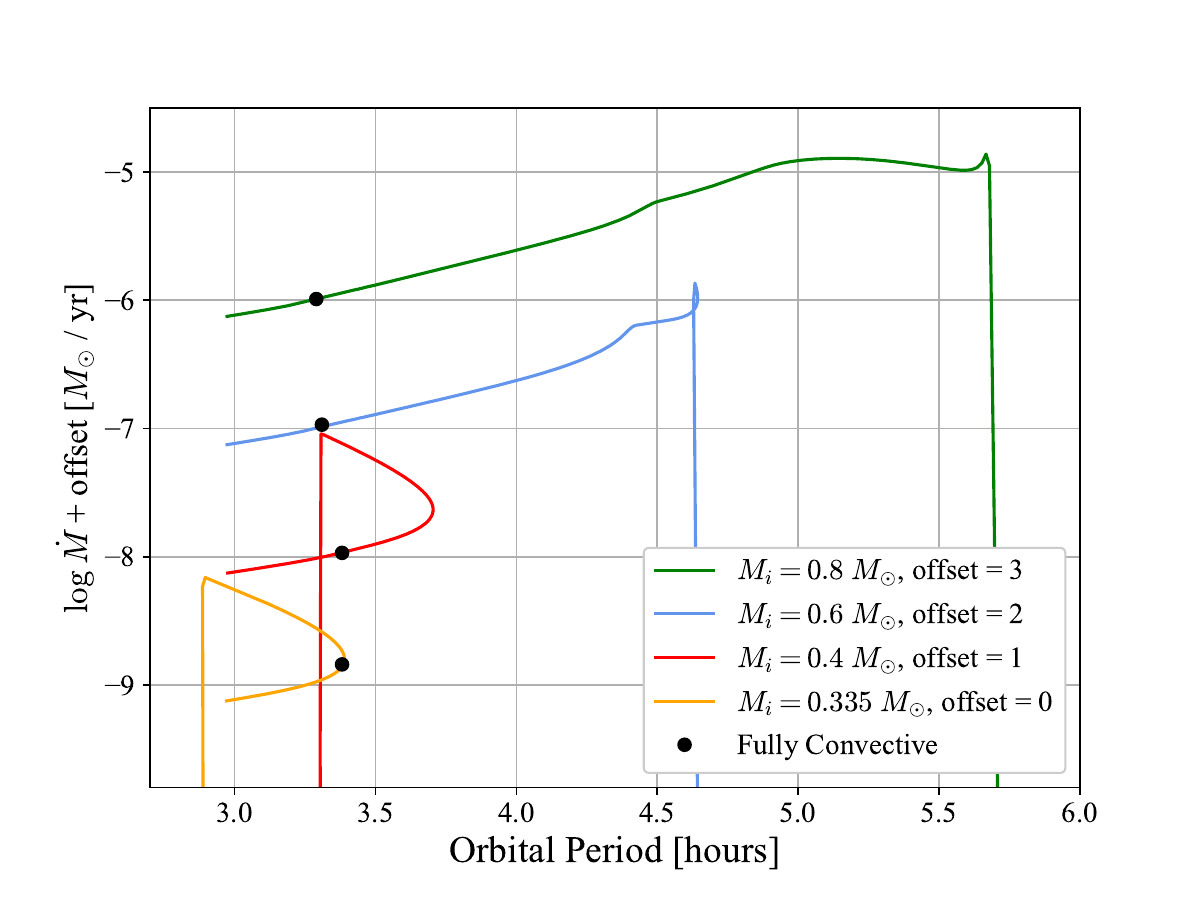}
    \caption{Mass loss rate vs. orbital period for models including magnetic braking. Each starting mass is given an offset in $\log\dot{M}$ for clarity. The black dots denote the transitions to fully convective. Each model starts at an orbital period of 9.5 hours.}
    \label{fig: m dot with MB}
\end{figure}

\begin{figure}
    \centering    
    \includegraphics[width=0.45\textwidth]{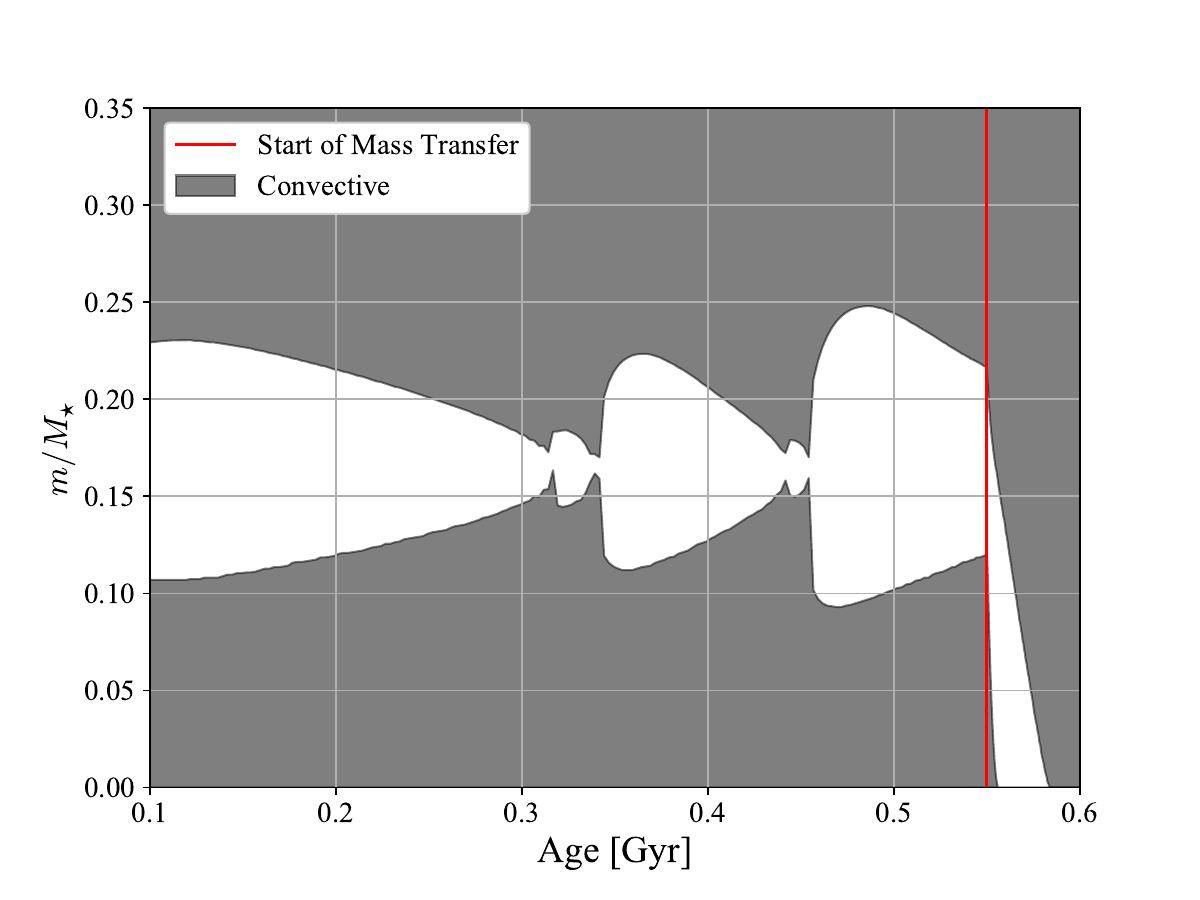}
    \caption{Mass coordinate of the convective \revI{(grey)} and radiative \revI{(white)} zones vs. age for a starting mass of 0.335 $M_{\odot}$ with magnetic braking. The vertical red line denotes the start of mass transfer.}
    \label{fig: conv zones with MB}
\end{figure}

Figure \ref{fig: m dot with MB} plots the mass loss rate vs. the orbital period for the four initial masses. The black points on the curves denote the transition to fully convective. We do not find any change in the mass loss rate when crossing these points, verifying that the transition is indeed smooth. Figure \ref{fig: conv zones with MB} displays the mass coordinate of the convective zones for the model with a starting mass of $0.335$ $M_{\odot}$. CKI cycles occur, but are abruptly halted once mass transfer begins. From these models we conclude that when magnetic braking acts via the \cite{RVJ} prescription with $\gamma = 3$, the CKI \revI{either does not occur or stops abruptly at the onset of mass transfer}. The CKI does not cause a detachment and cannot cause the period gap.

\subsection{Models without Magnetic Braking}

\begin{figure}
    \gridline{\fig{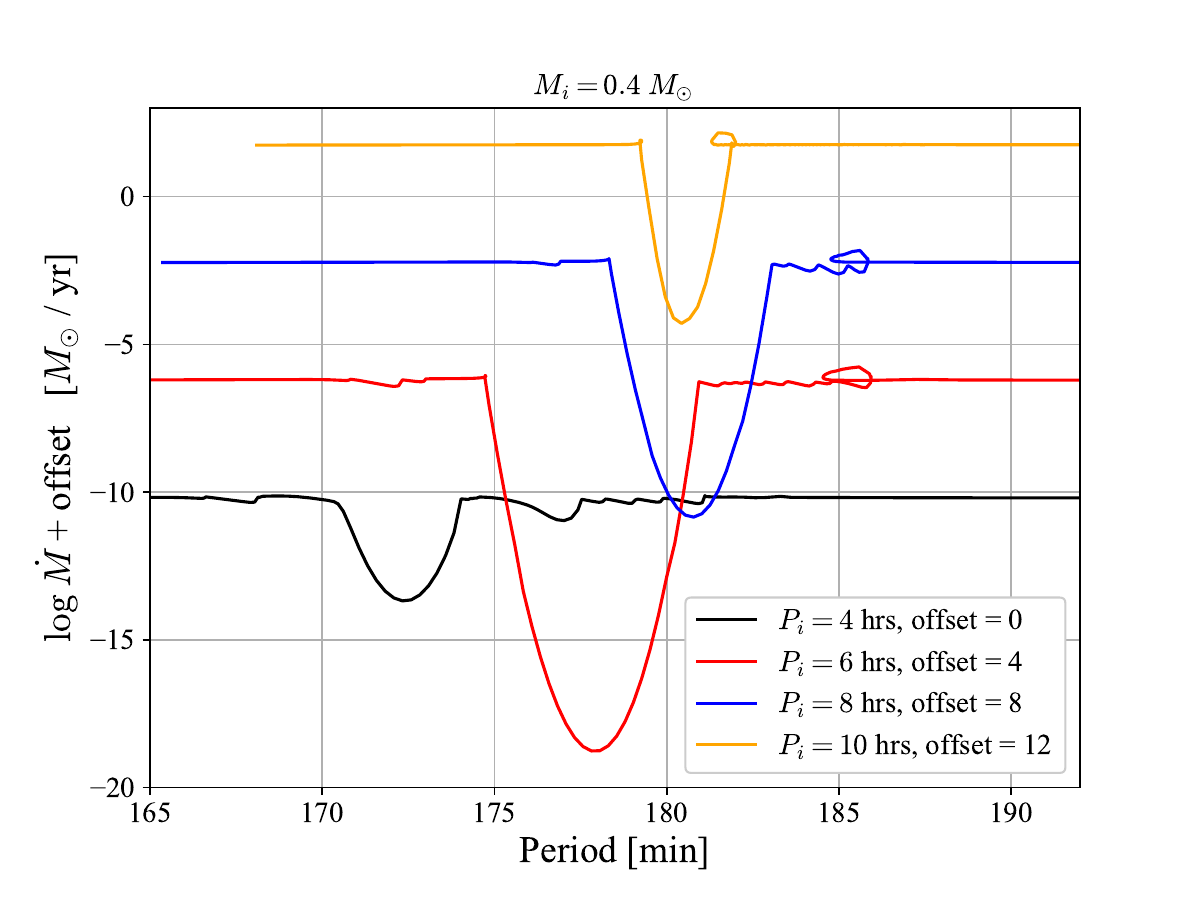}{\linewidth}{(a)}}
    \gridline{\fig{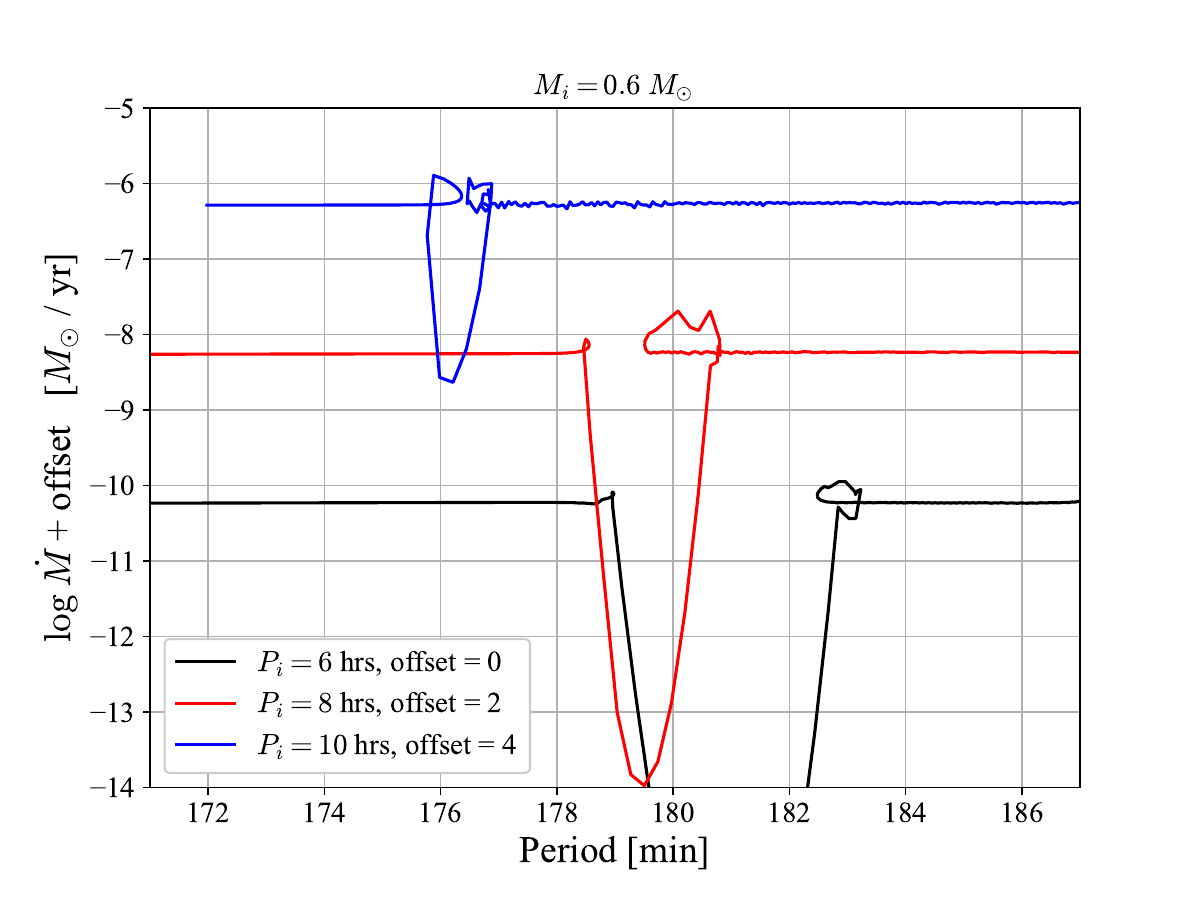}{\linewidth}{(b)}}
    \caption{Mass loss rate vs. orbital period for models including only gravitational radiation. Each starting period is given an offset in $\log \dot{M}$ for clarity. For (a) the starting mass is 0.4 $M_{\odot}$ and for (b) the starting mass is 0.6 $M_{\odot}$.}
    \label{fig: M dot v P only GR}
\end{figure}

When gravitational radiation is the sole source of AML, then the mass loss rates are low enough that the CKI can occur. To test the impact with lower mass loss rates, we computed models with a starting mass of 0.4 $M_{\odot}$ at starting periods of 4, 6, 8 and 10 hours. Figure \ref{fig: M dot v P only GR}a plots the mass loss rate vs. orbital period for the four initial periods. Each model displays one dip where the mass loss rate drops by several orders of magnitude. For the rest of the paper, we refer to this large reduction in mass loss rate as the \enquote{main dip}. These main dips cause period gaps. The \revI{width} of these gaps are small: 3.5 minutes for $P_{i} = 4$ hours, 6.1 minutes for $P_{i} = 6$ hours, 4.6 minutes for $P_{i} = 8$ hours and 2.0 minutes for $P_{i} = 10$ hours. These are all far smaller than the ${\sim}1$ hour period gap observed for CVs. In addition to the main dip, there are smaller dips in the mass loss rate. These smaller fluctuations become less apparent when the initial period increases. For the model with $P_{i} = 10$ hours, there are no other dips besides the main dip.

\begin{figure*}
    \gridline{\fig{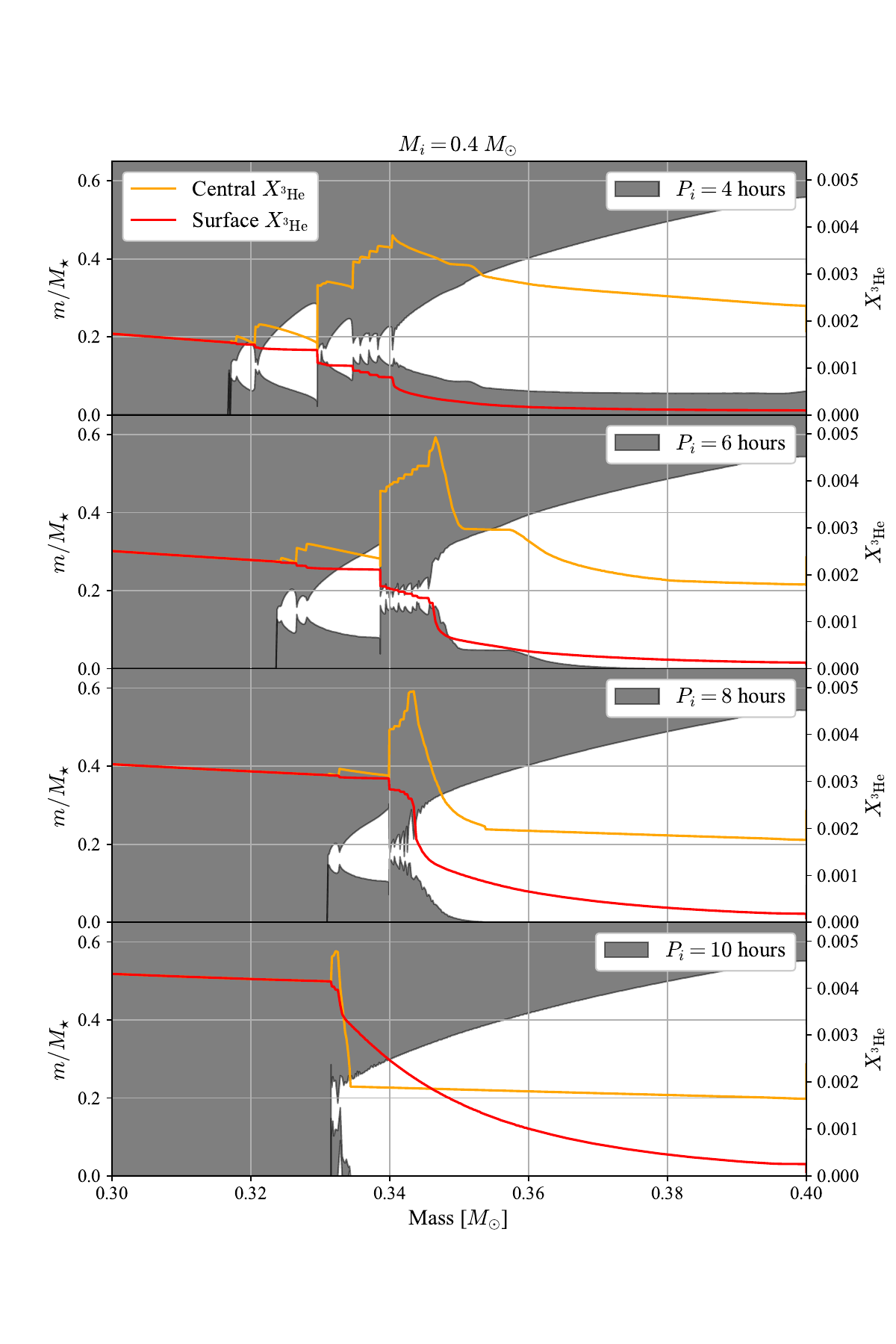}{0.5\textwidth}{(a)}
          \fig{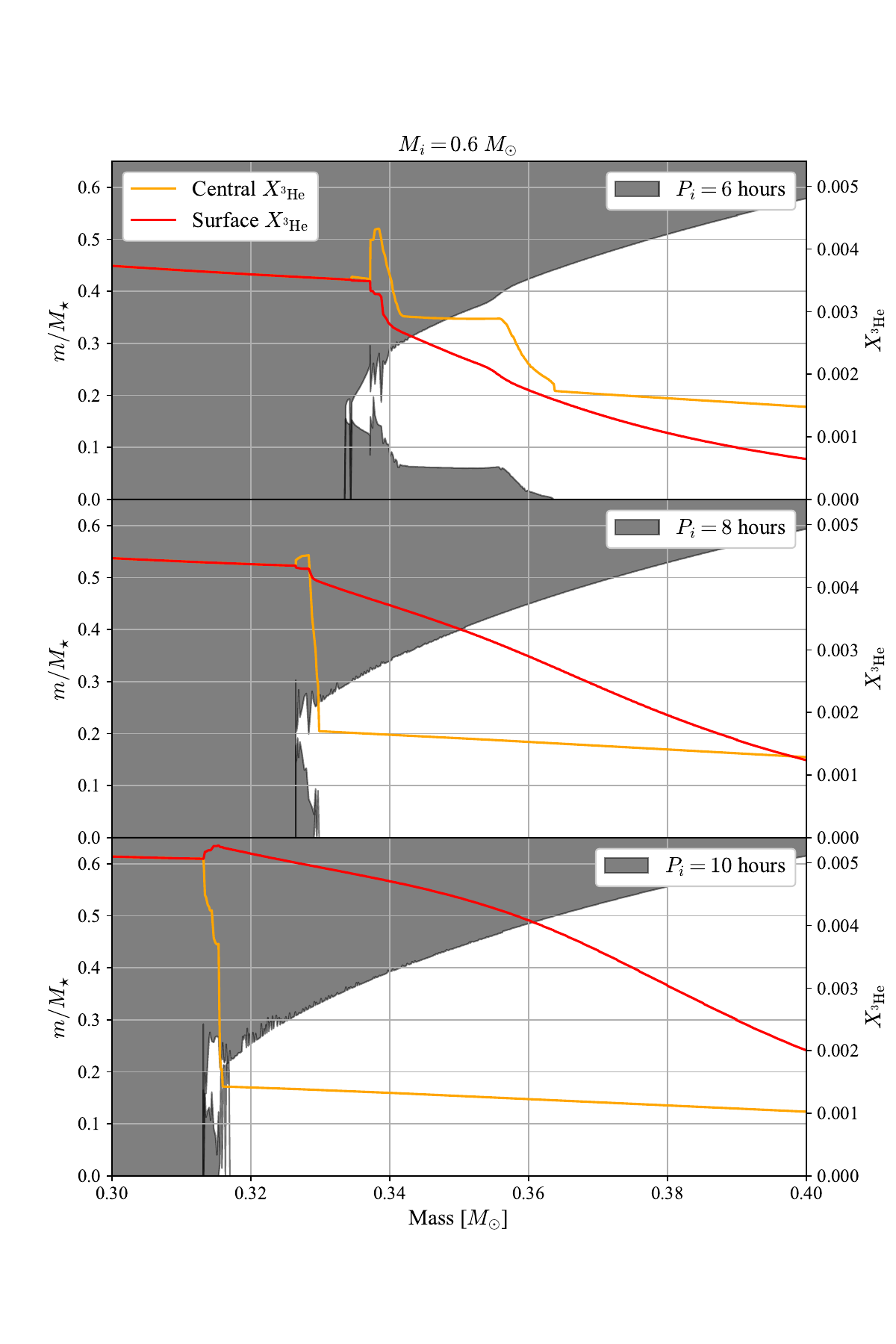}{0.5\textwidth}{(b)}}
    \caption{Mass coordinate of the convective zones (grey) and radiative zones (white) for the different starting periods. The orange/red curve plots the central/surface $^{3}$He mass fraction. The $x$-axis is the mass of the secondary. Since the stars are losing mass, the evolution in time is read from right to left. For (a) the starting mass is 0.4 $M_{\odot}$ and for (b) the starting mass is 0.6 $M_{\odot}$.}
    \label{fig: conv zones no MB}
\end{figure*}

Figure \ref{fig: conv zones no MB}a plots the mass coordinate of the convective and radiative zones along with the central and surface $^{3}$He mass fractions for the four starting periods. The $x$-axis in these plots is the secondary mass. Since the secondary loses mass, the evolution in time is read from right to left. When the core and envelope convection zones approach, the models undergo several small CKI cycles before a larger cycle equalizes the surface and central $X_{^{3}\mathrm{He}}$. This larger cycle causes the main dip in figure \ref{fig: M dot v P only GR}a. The strength of this cycle is controlled by the difference in the $^{3}$He mass fraction at the center and surface prior to beginning CKI cycles. When this difference is larger, the CKI pulsation and the main dip are larger. For the $P_{i} = 6$ hours model, the difference is the largest and thus the CKI pulsation and the main dip are largest. 

\begin{figure}
    \gridline{\fig{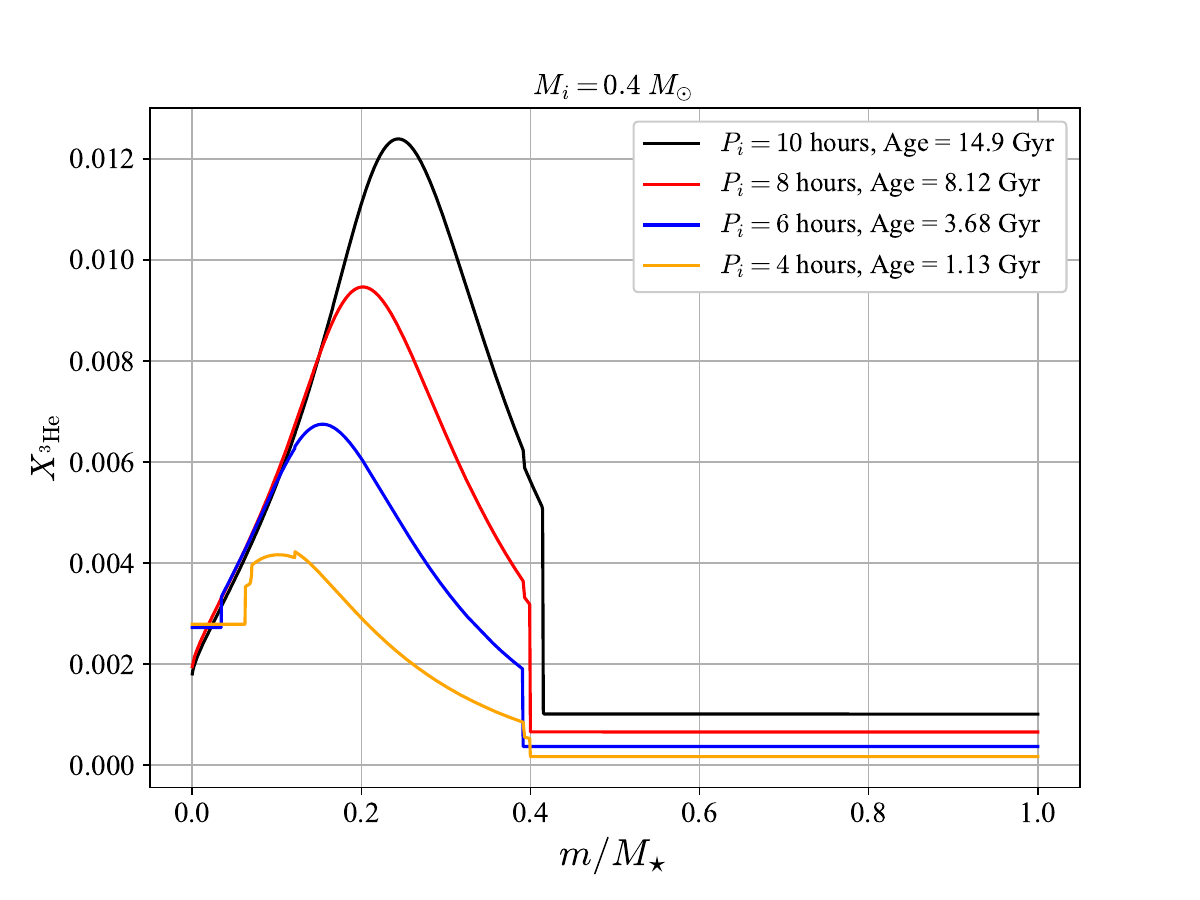}{\linewidth}{(a)}}
    \gridline{\fig{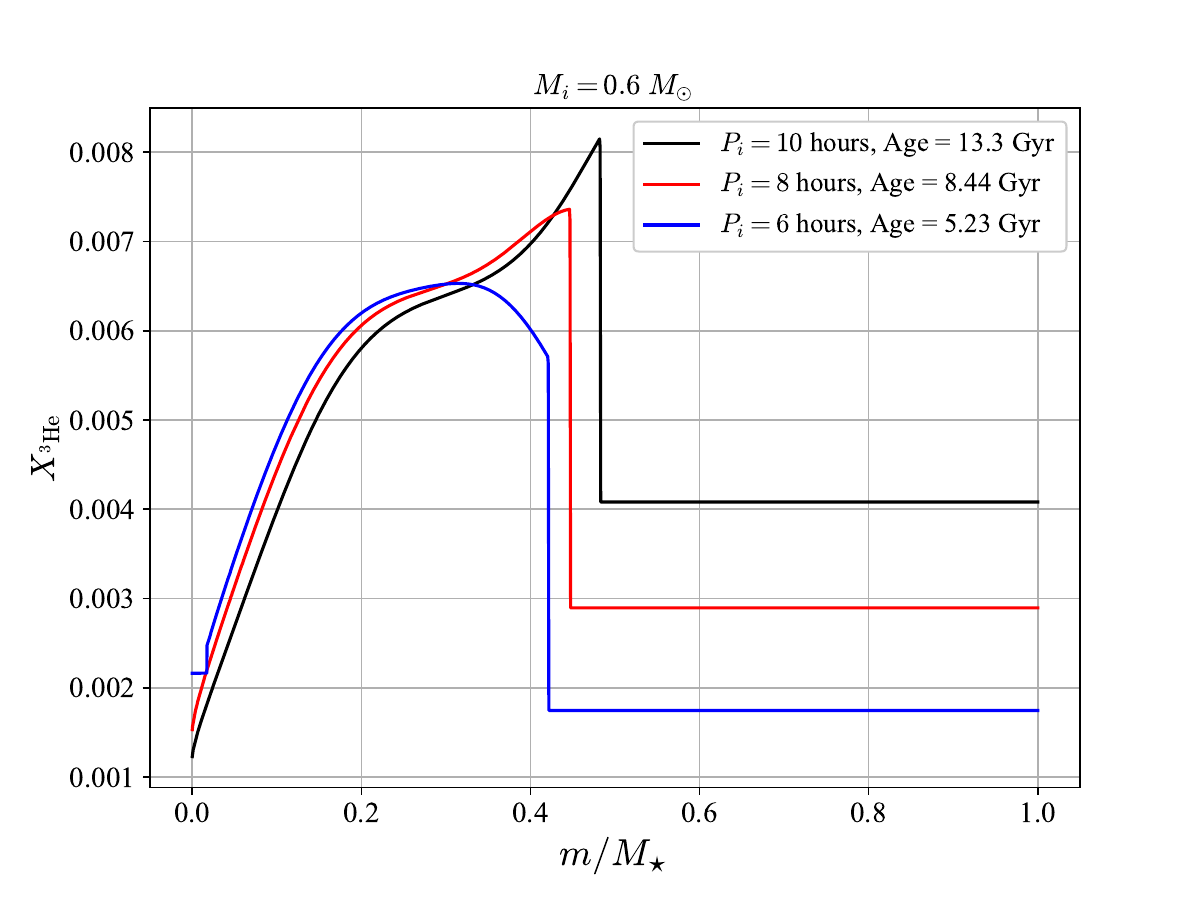}{\linewidth}{(b)}}
    \caption{Profiles of the $^{3}$He mass fraction for the different starting periods at a mass of 0.36 $M_{\odot}$. For (a) the starting mass is 0.4 $M_{\odot}$ and for (b) the starting mass is 0.6 $M_{\odot}$.}
    \label{fig: 3He profiles}
\end{figure}

We can understand this behavior by examining the distribution of $^{3}$He in the stars just prior to beginning CKI cycles. Figure \ref{fig: 3He profiles} plots the $^{3}$He mass fraction for each starting period when the mass is 0.36 $M_{\odot}$. For a 0.4 $M_{\odot}$ star, $^{3}$He peaks further out in the inner radiative region, where the temperature is low enough for $^{3}$He to be created. From figure \ref{fig: 3He profiles}, we find that the peak increases and shifts further out when the initial period is larger. This is simply caused by the larger initial period models spending more time in the detached phase, giving more time to produce more $^{3}$He. In figure \ref{fig: 3He profiles}, we list the age of the star at 0.36 $M_{\odot}$ which ranges from just over 1 Gyr for $P_{i} = 4$ hours to almost 15 Gyr for $P_{i} = 10$ hours.

The increasing peak size favors models with larger initial periods, as there is more $^{3}$He to drive CKI cycles. However, the shifting of the peak further out in the star favors smaller initial periods. For the four models, the outer convective zone reaches down to $m / M_{\star} \sim 0.2$, therefore any $^{3}$He beyond this point is swept to the surface. These two competing features lead to the $P_{i} = 6$ hours model having the largest CKI pulsation and the largest main dip. For the $P_{i} = 4$ hours model, less $^{3}$He is produced and thus the CKI cycles are smaller. For the $P_{i} = 8$ hours and $P_{i} = 10$ hours, much of the $^{3}$He is brought to the surface yielding a smaller difference in center and surface $X_{^{3}\mathrm{He}}$.

To test the impact of a larger starting mass, we computed models with a starting mass of 0.6 $M_{\odot}$ at starting periods of 6, 8 and 10 hours. Figure \ref{fig: M dot v P only GR}b plots the mass loss rate vs orbital period. As the initial period increases, the main dip becomes smaller, just as with the $M_{i} = 0.4$ $M_{\odot}$ case. However, for the same initial period, the main dip is smaller for the $M_{i} = 0.6$ $M_{\odot}$ case. In figure \ref{fig: conv zones no MB}b we plot the mass coordinate of the convective and radiative zones along with the central and surface $^{3}$He mass fractions. For equal initial periods, the difference between the central and surface $^{3}$He mass fractions prior to merger of the two convective zones is smaller in the $M_{i} = 0.6$ $M_{\odot}$ case, which causes the main dips to be smaller. This is a consequence of the higher central temperatures for higher mass stars, which leads to a \revI{buildup} of $^{3}$He further out in the star. Figure \ref{fig: 3He profiles}b plots the $^{3}$He profiles for the three starting periods when the mass of the star is 0.36 $M_{\odot}$. We find that the peaks in $^{3}$He occur further out in the star and thus more $^{3}$He is brought to the surface as the outer convective zone pushes inwards. An additional feature apparent in the $M_{i} = 0.6$ $M_{\odot}$ models, but also apparent in the $M_{i} = 0.4$ $M_{\odot}$ models, is the increase in mass loss rate prior to the main dip. This is caused by an influx of $^{3}$He into the center of the star as the inner convective zone reforms and moves outwards. This increase in $X_{^{3}\mathrm{He}}$ increases the nuclear luminosity, which increases the stellar radius and consequently increases the mass loss rate.

Of special note is the $M_{i} = 0.6$ $M_{\odot}$, $P_{i} = 10$ hours model. From figure \ref{fig: conv zones no MB}b we find that the surface $X_{^{3}\mathrm{He}}$ is larger than the central $X_{^{3}\mathrm{He}}$ when the convective zones merge, thus one would not expect the CKI to occur. However, a small detachment phase still occurs. This model is quite evolved, with a central hydrogen mass fraction of ${\sim}0.55$ just prior to the formation of the central convective zone. When the central convective zone develops and the two convective zones merge, the central mean molecular weight decreases as $^{1}$H and $^{3}$He are mixed inwards and $^{4}$He is mixed outwards. This drop in central mean molecular weight causes abrupt drops in central temperature. It is the competition between the decreasing central temperature, which decreases the nuclear luminosity, and the increasing central $X_{^{3}\mathrm{He}}$, which increases the nuclear luminosity, that controls the behavior. When the convective core first develops, the increase in central $^{3}$He outweighs the decreasing central temperature. The nuclear luminosity increases, causing the stellar radius to increase which in turn causes an increase in the mass loss rate. When the two convective zones eventually merge, the temperature drops low enough that the nuclear luminosity abruptly decreases, causing the radius to sharply drop which yields the small detachment period.

\section{Discussion} \label{sec:discussion}
We have presented models demonstrating the role of the CKI in CVs. In short, for \enquote{normal} CVs which evolve via magnetic braking above the period gap, the CKI has no effect. For models undergoing the CKI, the high mass loss rates driven by magnetic braking immediately halt CKI cycles. This is consistent with \revI{the findings of} \cite{CKI_initial} who demonstrated that for high mass loss rates which drive the star out of thermal equilibrium, the CKI does not occur. For models which do not naturally undergo the CKI, the high mass loss rates driven by magnetic braking cause the fully convective transition to occur at lower masses (${\sim}0.22$ $M_{\odot}$). At these low masses, the central temperatures are much lower than at the usual CKI mass range (${\sim}0.35$ $M_{\odot}$), therefore the CKI does not occur. This confirms the idea presented in \cite{Baraffe_Chabrier_2018}. 

When the systems evolve only via gravitational radiation, the \revI{secondary remains in thermal equilibrium and the} mass loss rates are small enough for the CKI to occur. The cycles cause a dip in the mass loss rate by several orders of magnitude, an event which we term the main dip for this paper. We find that for a given starting mass, the size of the main dip depends on the difference between the central and surface $^{3}$He mass fraction. For the four initial periods tested, the $P_{i} = 6$ hours model has the largest main dip. As the initial period increases from 6 hours, the main dip decreases in depth and length. We also test a larger initial mass of 0.6 $M_{\odot}$ and find that the main dip becomes smaller in depth and size compared to the 0.4 $M_{\odot}$ model at the same initial period. Generally, we find that increasing the initial period or increasing the initial mass leads to smaller main dips. The main dips are small, the largest are on the order of a few minutes. These are far smaller than the hour long period gap observed for CVs. Therefore, even in the idealistic case where angular momentum loss is driven only by gravitational radiation, the CKI cannot cause the CV period gap.

Ultimately, it does not seem that the CKI has any relevance for \enquote{normal} CVs. It has no effect when magnetic braking is active and if only gravitational radiation acts, the mass loss dips are far too small to cause the CV period gap. However, as there are still questions regarding the disrupted magnetic braking model, it is important to test new ideas with stellar evolution models. It is possible that the CKI may be relevant for CVs with a highly magnetic WD (i.e. polars). In polars, magnetic braking is believed to be weaker as part of the wind from the secondary becomes trapped in the WD magnetosphere \citep{WW02, Belloni_et_al_2019}. \cite{WW02} find that for WD magnetic moments $\gtrsim 4 \times 10^{34}$ G cm$^{3}$, the period gap disappears and mass transfer is driven primarily by gravitational radiation. The results in this report suggest that for these objects, an observable feature in parameter distributions may be present around $P_{i} = 3$ hours. It may be in the form of a small, few minute long period gap or possibly a greater variation in system-to-system mass loss rate. However, at this point not enough polars are known to test this prediction. In the PolarCat catalog \citep{polarcat}, there are only 71 objects with measured periods and field strengths and only 27 of these objects have orbital periods in the range 2 - 4 hours. Additionally, accurate masses/radii are often not available for polars, which, in addition to the field strength, is required to determine if the WD magnetic moment is large enough to suppress magnetic braking.

When gravitational radiation is the only AML mechanism, it takes a long time for the system to evolve through the detached phase. For both models with an initial period of 10 hours, the age of the stars when the mass drops to 0.36 $M_{\odot}$ is comparable to the age of the universe. The smaller initial period models, for $P_{i} = 4,6$ hours, the time taken to reach 0.36 $M_{\odot}$ is on the order of a few Gyr. These smaller initial period models produce larger main dips which have a better chance of producing an observable feature with a large enough sample. 

\revII{It should be noted that even strong field polars may be unable to completely suppress magnetic braking at the larger periods tested here. The required WD magnetic moment for complete magnetic braking suppression depends on the secondary mass which decreases with time due to mass transfer \citep[see][]{WW02}. A more rigorous treatment of the angular momentum loss would use a reduced magnetic braking prescription \citep{WW02, Belloni_et_al_2019}. In this work, we use the starting period to change the age of the star rather than as a true starting condition. By increasing the initial period, we can test the impact of the system spending a longer time before reaching the CV phase.} 

\revII{For polars, recent formation theories suggest that} the time taken in the detached phase may be more consistent with the smaller initial period models. \cite{Cryst_dynamo} have presented a formation mechanism for polars where the WD magnetic field is generated by a crystallization-rotation dynamo. The evolution occurs in six stages. The initial detached phase is spent as a non-magnetic pre-CV where AML is driven by gravitational radiation and magnetic braking (stage 1). Once mass transfer is initiated, accretion spins up the WD (stage 2). If the core of the WD has begun crystallizing, the spin up yields the appropriate conditions for a crystallization-rotation dynamo (stage 3). If the WD magnetic field is strong enough to connect with the field of the secondary, the WD becomes synchronized, which transfers spin angular momentum into orbital angular momentum (stage 4). The orbital period increases and the system becomes detached again. Once synchronization is complete, the orbital period decreases through weakened magnetic braking or, if the WD field is strong enough, through only gravitational radiation (stage 5). Eventually mass transfer resumes forming a polar (stage 6). \cite{Time_dependent_WD_field} have demonstrated that this scenario does not require the crystallization-rotation dynamo and can be reproduced by assuming the emergence of the field at the surface of the WD is a function of age.

If this scenario is correct, then the long detached phase (stage 1) occurs much quicker as magnetic braking provides a strong AML source. Our models with only gravitational radiation with the smallest initial periods seem better in agreement with the evolutionary sequence of \cite{Cryst_dynamo}. These models essentially start at stage 5, when the strong WD field has been generated and the system is evolving to mass transfer via gravitational radiation. Therefore, while the main dips become smaller as the initial orbital period increases, new theories on the formation of polars hint that the time until mass transfer is more in agreement with our smaller initial period models.

\section{Conclusion} \label{sec:conclusion}
We have presented stellar evolution models using \texttt{MESA} demonstrating the role of the CKI on CV evolution. For \enquote{normal} CVs which evolve with magnetic braking and gravitational radiation above the period gap, the CKI has no effect as CKI cycles either do not occur or are abruptly halted once mass transfer begins. If the only AML mechanism is gravitational radiation, then the CKI does occur which can cause a small detachment phase resulting in a small period gap \revI{with a width of} a few minutes. This may be relevant for the evolution of high field polars, where the WD magnetic field is strong enough to suppress magnetic braking. Our results indicate that the CKI can cause a small detachment phase in these objects at an orbital period around 3 hours. Without the CKI, no period gap is predicted for these objects. Currently, the number of these objects is small. As the number of well studied polars increases, one can search for an observable signature of the CKI acting in the distribution of strong field polars.

C.L. gratefully acknowledges support from the Delaware Space Grant College and Fellowship Program (NASA Grant 80NSSC20M0045).

\bibliography{Bibliography}{}
\bibliographystyle{aasjournal}

\end{document}